# Diverse surface waves supported by bianisotropic metasurfaces

YOU Ou-bo[1], GAO Wen-long[4], LIU Ya-chao[5], XIANG Yuan-jiang[3*], ZHANG Shuang[1,2*]

1. Department of Physics, The University of Hong Kong, Hong Kong, China
2. Department of Electrical & Electronic Engineering, The University of Hong Kong, China
3. School of Physics and Electronics, Hunan University, Changsha 410082, China
4. Department of Physics, Paderborn University, Warburger Straße 100, 33098 Paderborn, Germany.
5. Institute of Microscale Optoelectronics, Shenzhen University, Shenzhen 518060, China

*Corresponding authors: XIANG Yuan-jiang: xiangyuanjiang@126.com; ZHANG Shuang: shuzhang@hku.hk;

**Abstract:** Surface waves supported by structured metallic surfaces, i.e. metasurfaces, have drawn wide attention recently. They are promising for various applications ranging from integrated photonic circuits to imaging and bio-sensing in various frequency regimes. In this work, we show that surface states with diverse polarization configurations can be supported by a metasurface consisting of a single layer of bianisotropic metamaterial elements. The structure possesses $D_{2d}$ symmetry, which includes mirror symmetry in the $xz$ and $yz$ plane, and $C_2$ rotational symmetry along $y = \pm x$ axis. Due to this unique symmetry, the metasuface supports both transverse electric (TE) and transverse magnetic (TM) waves along $k_x$ and $k_y$ directions, while a purely longitudinal mode and an elliptically polarized transverse electromagnetic (TEM) mode along $k_y = \pm k_x$ directions. The versatility of the surface modes on the metasurface may lead to new surface wave phenomena and device applications.

Surface plasmon polaritons (SPPs), due to their tight confinement to a metal/dielectric interface and large wave vectors, represent an important platform for various applications ranging from integrated photonic circuits to sensing applications [1, 2]. However, at optical frequencies, due to the significant ohmic loss of metals, the applications of SPPs suffer from the short propagation lengths. At longer wavelengths, the loss of SPPs is significantly reduced at the cost of poor confinement of the SPPs in the dielectric material. In 2004, Pendry proposed the concept of spoof plasmon that could be supported by a corrugated metallic surface with an effective plasma frequency determined by the geometries of the metallic structures [3], which was subsequently experimentally verified [4]. This new scheme greatly improves the confinement of surface waves to the structured surfaces, and has attracted tremendous interests from the community of photonics. Various explorations have been carried out based on spoof plasmons, including rainbow slow light trapping effect [5, 6], focusing of terahertz waves [7], terahertz subwavelength waveguides [8], and terahertz sensing [9].

Indeed, structured surfaces (or metausrfaces) can be engineered to provide more diverse functionalities that go beyond confinement of surface waves, such as wavefront and amplitude control [10-18], enhanced and tailored nonlinear optical processes [19-23], resulting in a wide range of applications including imaging, holography and bio-sensing [24-30]. These new functionabilities can arise from judicious engineering of the unit cells, benefitting from the unconventional electromagnetic responses of complex metamaterial designs such as artificial magnetism, hyperbolicity, chirality and bianisotropy [31-38]. Bianisotropy refers to a cross coupling between electric and magnetic responses along orthogonal directions. It can exist in structures that lack inversion symmetry but with preserved mirror symmetry. Bianisotripic metamaterials have shown some highly intriguing phenomena such as asymmetric absorption [39, 40], optical spin-orbit coupling [41], and topological optical effects [42]. In the past decade, biansotropy has been employed for designing topological metamaterials, which have shown interesting phenomena such as Fermi arc states and transverse spin of bulk optical modes [43-46]. Compared to three dimensional bulk metamaterials, the ultrathin nature of bianisotropic metasurfaces could lead to more practical applications due to its low fabrication cost and highly compact physical sizes.

In this work, we experimentally investigate the surface states supported by a bianisotropic metasurface and showcase a number of interesting effects – the existence of both TE and TM surface modes along certain directions [47], while helical transverse electromagnetic mode and longitudinal mode in some other directions. The configuration of the metasurface is illustrated in Fig. 1(a). Each unit cell of the metasurface consists of a saddle-shaped metallic loop. The same unit cell, when arranged in a three dimensional array, forms a type-I ideal Weyl metamaterial, as demonstrated previously [43]. Here we are interested in a metasurface consisting of a single layer of such structure, and therefore the bulk property is not well defined. Each unit cell can be considered as two perpendicular split ring resonators with opposite orientation of the openings. The structure possesses $D_{2d}$ symmetry, which includes mirror symmetry in the *xz* and *yz* plane, and $C_2$ rotational symmetry along $y = \pm x$ axis. The fundamental resonant mode of the unit cell is a combination of electrical dipole moment and magnetic dipole moment, each of which can be excited by both an electric field or a magnetic field oriented in the *x-y* plane.

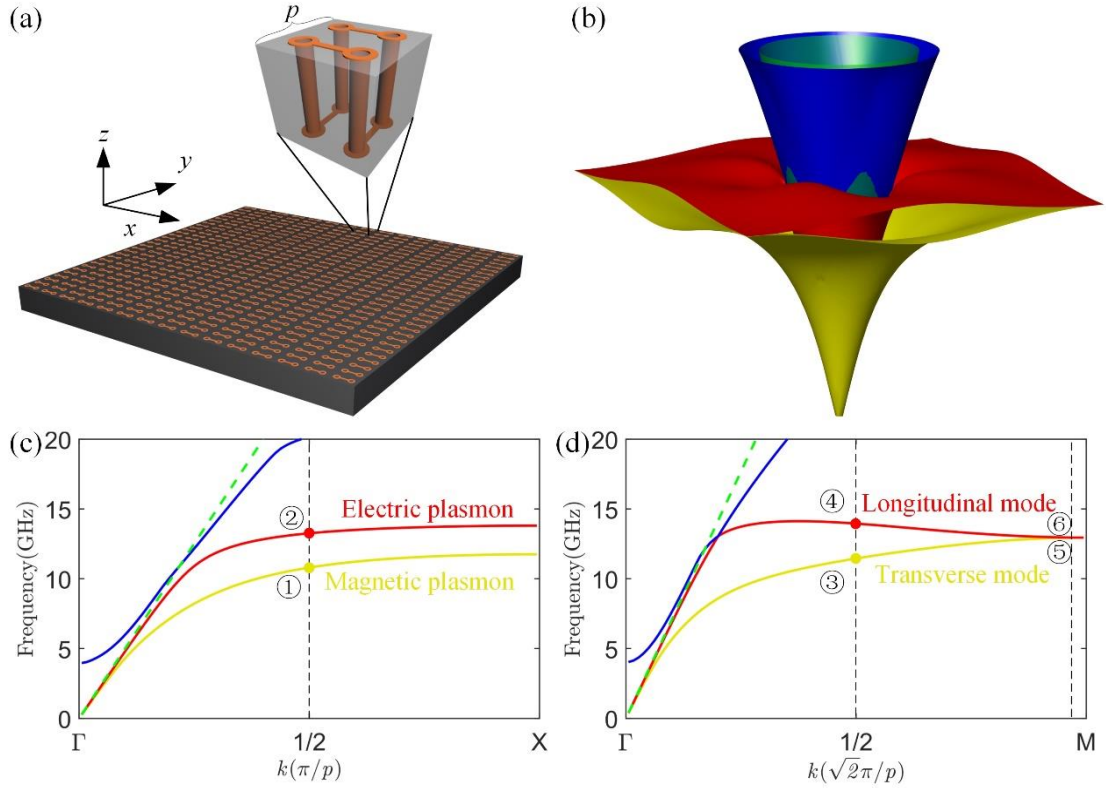

Fig.1 (a) The schematic of the single-layer metasurface. Every unit cell consists of a saddle-shaped metallic inclusion possessing $D_{2d}$ point symmetry embedded in the dielectric substrate whose relative permittivity is 2.2. The period of the metasurface along $k_x$ or $k_y$ is $p$. (b) The band structure of the metasurface. The 1st, 2nd, 3rd bands and lightcone are plotted in yellow, red, blue and green, respectively. (c) The dispersions of the surface modes along $k_y=0$ direction. The modes along this direction corresponding to the 1st and 2nd bands can be regarded as Electric plasmon and Magnetic plasmon, respectively. (d) The dispersions of the surface modes along $k_y=k_x$ direction. The modes along this direction for the 1st and 2nd bands can be regarded as transverse mode (elliptically polarized) and longitudinal mode, respectively.

The band structure of the metasurface is shown in Fig. 1(b). There exist multiple bands in the system, with some bands located very close to the light cone (for example, the blue one). Here we are interested in the 1st and 2nd modes (yellow and red) with larger wave numbers. Due to the $D_{2d}$ symmetry of the system, the wave propagation along **x** and **y** directions can be related to each other by simply a rotation of π about the $k_x=\pm k_y$ axis. In order to have a clearer view of the dispersions of the surface waves, we plot the dispersions along the $k_x/k_y$ direction in Fig. 1(c). The two lowest modes, labelled electric and magnetic plasmon in the figure, exhibit the typical surface plasmon dispersion features, i.e. an approximately linear dispersion at lower frequency, and the dispersion gradually become flat when approaching the effective plasma frequency [3]. We further plot the dispersions of the surface modes along $k_x=\pm k_y$ directions, as shown in Fig. 1(d). These 1st and 2nd modes are TEM and longitudinal modes, respectively. Interestingly, these two surface modes become degenerate at the corner of the Brillounin zone, i.e. **M** point. As will be explained later, this degeneracy arises from the $D_{2d}$ symmetry of the system.

The existence of TE and TM modes can be analyzed through the point group symmetry. For modes along $k_x/k_y$ direction, they must satisfy the mirror symmetry $M_{x/y}$ about *yz* or *xz* plane. The eigenvalues of M are $\pm 1$. For M = +1, the normal **E** field with respect to mirror plane will cancel out when integrated over the unit cell, while the parallel **E** field remains. Meanwhile, the **H** field, which is a pseudo vector field, behaves in the opposite way. Therefore this represents a TM mode. The M = -1 mode, on the other hand, represents a TE mode based on a similar analysis. We further carry out full wave simulation, and present the field plots of the *x*-propagating modes in the *xz* cross-section plane cutting through the center of the unit cell in Fig. 2. For point ① on 1st mode in Fig. 1(c), the electric field is aligned along **y** direction, which is perpendicular to the propagation plane (Fig. 2(a)), whereas the magnetic field lies in the propagation plane having both *x* and *z* components (Fig. 2(b)). This confirms that 1st mode is not just a TE mode but also a magnetic surface plasmon mode, which is distinct from the conventional surface plasmon mode. It is interesting to note that the electric and magnetic field components are mostly confined to the top and bottom surfaces, respectively. On the other hand, the field distributions (Fig. 2(c) and 2(d)) of point ② on 2nd mode show opposite configuration as that of TE mode. Namely, the electric field lies in the propagation plane and the magnetic field is perpendicular to it, which represent the main features of conventional TM polarized surface plasmon modes. The presence of both TE and TM polarized surface plasmon modes can be attributed to the fact that the unit cell of the bianisotropic metasurface supports both electric and magnetic resonances.

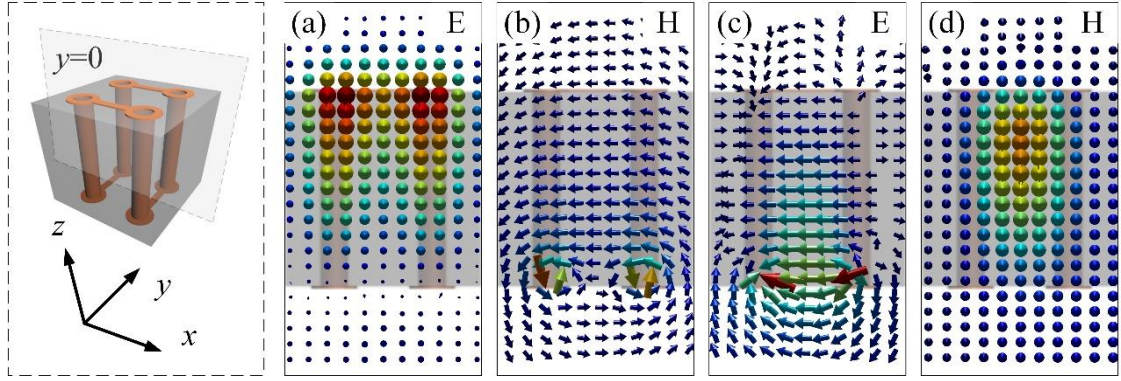

Fig. 2 Field distributions of surface plasmon modes propagating along **x** direction. On the left, the schematic of the unit cell illustrating the plane in which the fields are plotted is shown. (a, b) The **E** and **H** field distributions for point ① on 1st mode in Fig. 1(c). (c, d) The **E** and **H** fields for point ② on 2nd mode. $k_x$ of point ① and ② is fixed at $\pi/2p$, where $p$ is the period along **x** and **y** directions.

However, away from $k_x/k_y$ directions, the surface plasmon modes are not exactly TE and TM modes anymore, but generally a hybridization between them. Interestingly, along $k_x=\pm k_y$ directions, this hybridization leads to a complete re-arrangement of the field components, and the longitudinal mode and TEM mode emerge. The existence of the longitudinal mode and TEM can also be analyzed through the symmetry of the point group with respect to this direction. As the metasurface has $C_2$ symmetry along $k_x=\pm k_y$ directions, the eigenvalues of $C_2$ are $\pm 1$. For the $C_2=+1$ mode, if we rotate the field by $\pi$ about the symmetry axis, we will get the same field. This means that all the field components, both **E** and **H** perpendicular to the symmetry axis will cancel out when integrated over the unit cell, while fields components parallel to the symmetry axis remains, and therefore this corresponds to a longitudinal mode. For the $C_2=-1$ mode, the situation is opposite, i.e. all longitudinal components cancel out while transverse components remain, which corresponds to a TEM mode. This is illustrated by the fields shown in Fig. 3. Fig. 3(a) and 3(b) show the distribution of the electric and magnetic fields of point ③ ($k_x=k_y=\pi/2p$) in Fig. 1(d), respectively, in a cross section plane perpendicular to the propagation direction. It is observed that both the **E** and **H** fields primarily lie in the plane, while the longitudinal components of the fields at different locations are opposite and cancel out, leading to an overall TEM mode. It is interesting to note that both the **E** and **H** fields rotate anticlockwise with time in the plane, i.e. the TEM mode is elliptically polarized. On the other hand, the distribution of **E** and **H** fields of point ④, as illustrated in Fig. 3(c) and 3(d) respectively, are primarily aligned along the propagation direction. Thus, it is confirmed that 1$^{st}$ mode is a pure longitudinal mode with both longitudinal **E** and longitudinal **H** components. We further look into the field distributions of the two points ⑤ and ⑥ in Fig. 1(d) close to **M** point in a horizontal plane (*xy* plane) cutting through the center of the unit cell, as shown by Fig. 3(e-h). The fields clearly show that the two modes can be related to each other through the following symmetry operations: a rotation of 90° in the *xy* plane about the center of the unit cell, followed by a mirror symmetry in **z** direction, which are consistent with the $D_{2d}$ symmetry of the metasurface structure. Thus, this symmetry argument explains the degeneracy between the two modes at **M** point as shown in Fig. 1(d).

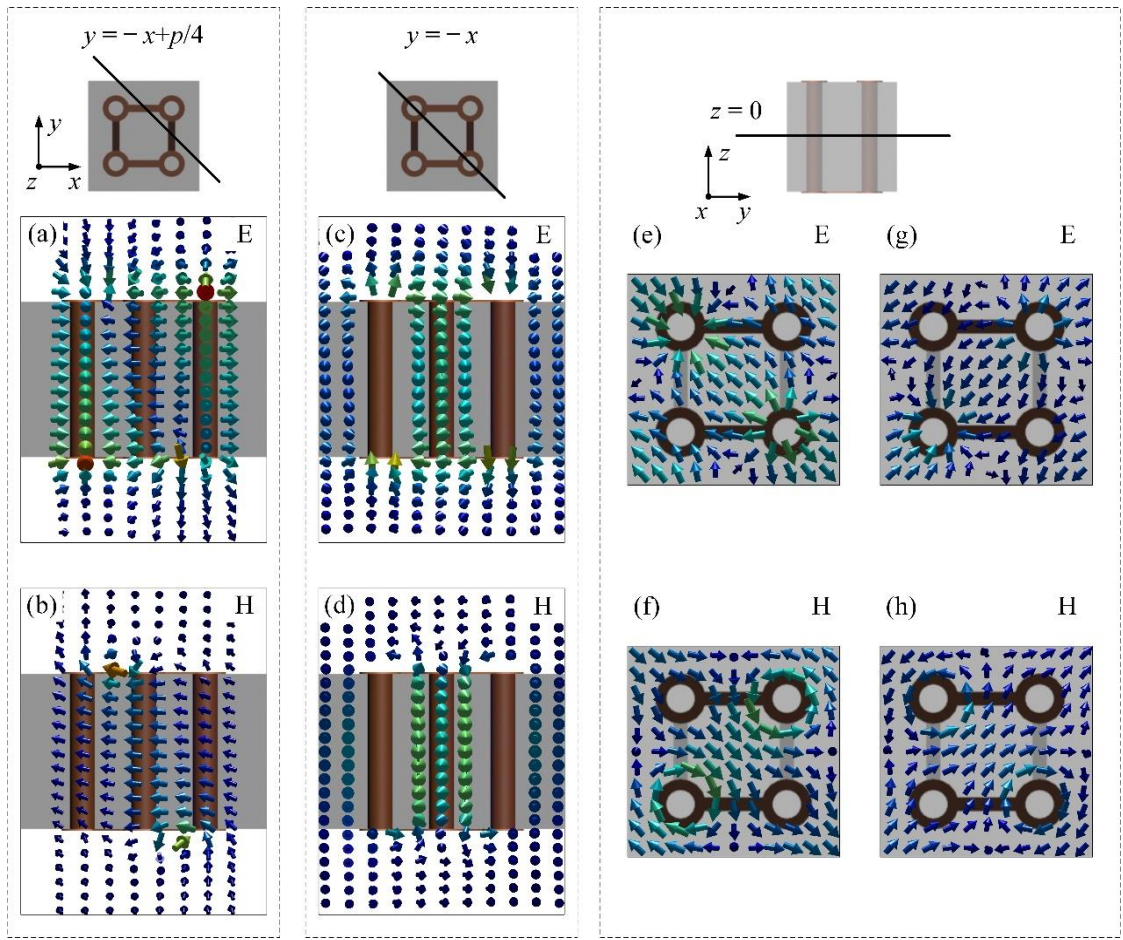

Fig. 3 Field distributions of surface plasmon modes propagating along $k_x = k_y$ direction. (a, b) Field distributions of 1st mode at point ③ ($k_x = k_y = \pi/2p$) in Fig. 1(d). The simulated **E** (a) and **H** (b) field distributions in the plane perpendicular to the propagation direction, corresponding to the cutting plane shown in the schematic above. In both plots, the overall field distribution lie in the plane, indicating that this is a TEM mode. (c, d) Field distributions of 2nd mode at point ④ ($k_x = k_y = \pi/2p$) in Fig. 1(d). The simulated **E** (c) and **H** (d) field distributions in the plane perpendicular to the propagation direction, corresponding to the cutting plane shown in the schematic above. In both plots, the overall field distributions are out of plane (along the propagation direction), indicating that this is a pure longitudinal mode with both longitudinal components of **E** and **H** fields. (e-h) Field distributions of points ⑤ and ⑥ in Fig. 1(d), closing to the **M** point (phase advance is 170°), for a horizontal $xy$ plane cutting through the center of the unit cell, as indicated by the schematic above. (e, f) correspond to the **E** and **H** field distributions of point ⑤, and (g, h) correspond to **E** and **H** field distributions of point ⑥. It is observed that the two modes are related to each other through an in-plane rotation of 90° about the center of the unit cell, followed by a mirror symmetry in **z** direction.

To measure the dispersion of the surface modes, we place a source antenna at the center of bottom surface of the sample, which consists of 90×70 unit cells, while the electric field distribution is mapped by a probe antenna raster-scanning the top surface. The Fourier transformations of the electric field, which represent the equal frequency contours (EFCs), at two representative frequencies of 10.9 GHz and 13.3 GHz are shown in Fig. 4(a) and 4(c), respectively, to illustrate the 1st and 2nd modes. The corresponding simulated results are shown in Fig. 4(b) and 4(d). The EFC of 1st band appears roughly as a round loop, as shown in Fig. 4(a, b). Inside the EFC of 1st band, the light cone and higher modes are crowded together into a bright smaller circle. The measured EFCs match well with the simulated ones shown in Fig. 4(b). At a higher frequency of 13.3 GHz, the EFC of 2nd band shows a more complicated pattern - an ellipse centered at the Γ point and four nearly straight lines close to the corner (Fig. 4(c)). Considering the periodic boundary of the Brilloun zone, these four lines indeed form a closed contour around the **M** point. The measured EFC agrees well with the simulation result (Fig. 4(d)), except for the missing of half of the elliptical contour with long axis oriented along **x** direction. This is because in the experiment only the top surface is measured, whereas the mode corresponding to the missing contour is mainly localized at the bottom surface. From the measured EFCs at different frequencies, one can retrieve the dispersion curves along different directions. As shown in Fig. 4(e), the experimentally retrieved dispersion of 1$^{st}$ and 2$^{nd}$ bands along $k_x/k_y$ directions clearly show the characteristics of typical surface plasmons and they correspond to the TE and TM surface plasmon modes with different effective electric and magnetic plasma frequencies. However, along $k_x=\pm k_y$ directions, the two bands show very distinct features – while mode 1 shows similar dispersion as a conventional surface plasmon, mode 2 exhibits a negative dispersion at large wavevectors (Fig. 4(f)). They become degenerate at **M** point, matching very well with the numerical results as indicated by the dashed lines.

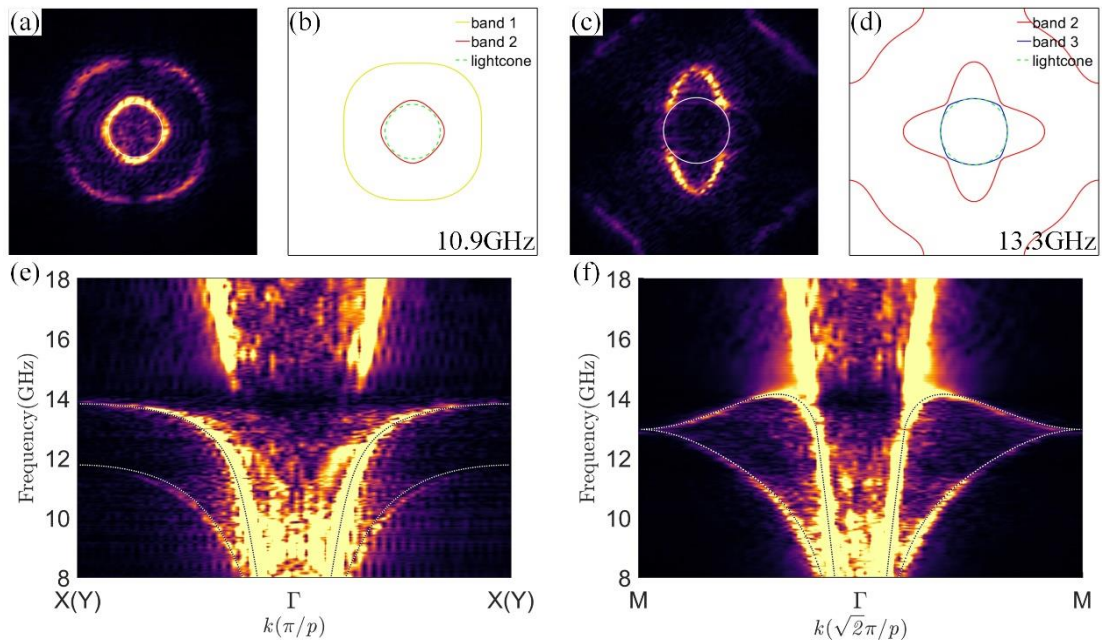

Fig. 4 Measured EFC and dipersion curves of the surface modes. (a, b) The measured (a) and simulated (b) EFC at frequency of 10.9 GHz. (c, d). The measured (c) and simulated (d) EFC at frequency of 13.3 GHz. (e, f) the dispersion of the surface modes along $k_x/k_y$ and $k_x=\pm k_y$ directions, respectively. The dashed lines in the plots correspond to the simulation results.

Finally we experimentally investigate the excitation of the surface waves by controlling the orientation of the source dipole antenna. The experimental setup for the measurement is shown in Fig. 5(a). A source antenna is oriented along either **x** or **z** direction in the middle of the edge along **x** direction. For both configurations, we measure the field distributions on either the top surface or the bottom surface of the metasurface. By combining the two measured field distributions with source dipole antenna oriented along the two orthogonal directions (**x** and **z**), one can retrieve the field distribution for tilted dipole antenna (e.g. orientation of +45° and -45°) and for circularly polarized antenna (left and right handed). Fig. 5(b) and 5(c) show the field patterns excited by a dipole antenna oriented along +45° and -45°, respectively, wherein the surface wave primarily propagates towards the left or the right hand side depending on the polarization of the exciting antenna. The field distributions also show very distinct features on the two sides when the source antenna is circularly polarized - a single beam appearing on one side and two split beams appearing on the other side, as shown by Fig. 5(d) and 5(e). The configurations are swapped when the rotating direction of the source antenna is flipped. This directly demonstrates the spin and orientation controlled excitation of the surface waves on the metasurface. In this experiment, the excitation efficiency is not very high, but sufficient to see all exotic features of this metasurface. For achieving a higher excitation efficiency, the size and orientation of the antenna would require very fine adjustment to match the polarization of the mode.

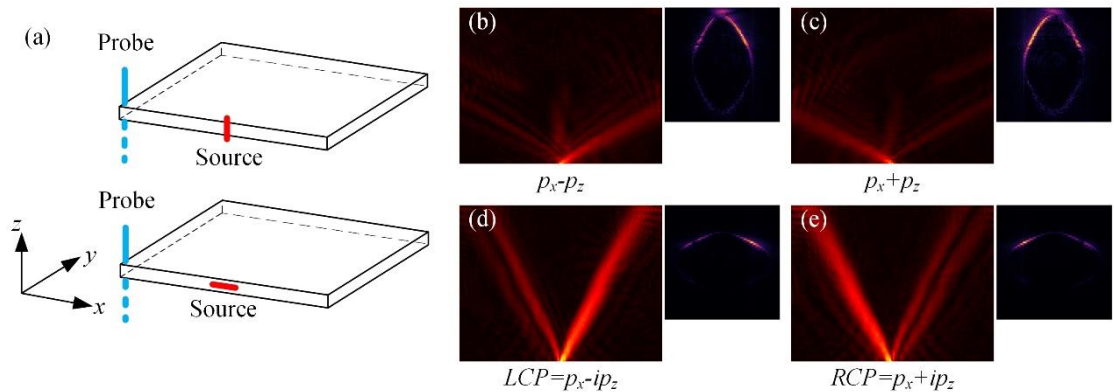

Fig. 5 Measurement of polarization controllable excitation of surface modes. (a) The experimental setup for measuring the surface mode. The excitation dipole antenna is oriented either in the vertical direction (upper panel) or the horizontal direction (lower panel) (b, c) Electric field distribution on top surfaces under polarization of $p_x-p_z$ and $p_x+p_z$, respectively. (d, e) same as (b, c) but the fields are measured on bottom surfaces under LCP and RCP excitation, respectively. All subplots attached to (b-e) are the corresponding EFCs in the Brillouin zone.

In summary, we have designed and demonstrated a bianisotropic metasurface with a unique symmetry configuration and investigated the rich features of surface waves supported by the metasurface. We have shown that both TE and TM surface plasmon waves can exist along certain directions, while along some other directions, there exist a pure longitudinal mode with both electric and magnetic components, and an elliptically polarized transverse electromagnetic mode. Such diverse dispersion and polarization configurations of the surface plasmon modes provide new degrees of freedom for constructing compact photonic integrated devices.

**Acknowledgments:**
Funding: This work was supported by the Research Grants Council of Hong Kong (AoE/P-502/20).